\documentclass{article}%
\usepackage{amsmath}
\usepackage{graphicx}%
\usepackage{amsfonts}%
\usepackage{amssymb}
\begin{document}

\title{Detailed discussion of a linear electric field frequency shift (important for
next generation) electric dipole moment searches ) induced in confined gases
by a magnetic field gradient: Implications for electric dipole moment
experiments (II)}
\author{A.L. Barabanov$^{\#}$, R. Golub$^{+}$ and SK Lamoreaux$^{\ast}$\\$^{\#}$Kurchatov Institute\\\ 123182 Moscow, Russia\\$^{+}$Physics Department\\\ North Carolina State University\\\ Raleigh, NC 27606\\$^{\ast}$University of California,\\\ Los Alamos National Laboratory\\\ Physics Division\\\ Los Alamos NM 87545}
\date{November 29, 2005}
\maketitle

\begin{abstract}
The search for particle electric dipole moments represents a most promising
way to search for physics beyond the standard model. A number of groups are
planning a new generation of experiments using stored gases of various kinds.
In order to achieve the target sensitivities it will be necessary to deal with
the systematic error resulting from the interaction of the well-known
$\overrightarrow{v}\times\overrightarrow{E}$ field with magnetic field
gradients (often referred to as the geometric phase effect \cite{cumm},
\cite{jmp}. This interaction produces a frequency shift linear in the electric
field, mimicking an edm. In this work we introduce an analytic model for the
correlation function which determines the behavior of the frequency shift
\cite{LG} and show in detail how it depends on the operating conditions of the
experiment. We also propose a method to directly measure ths correlation
function under the exact conditions of a given experiment.

\end{abstract}
\tableofcontents

\section{Introduction}

The proposition that the search for particle electric dipole moments (edm)
represents a reasonable method to \ look for physics beyond the standard model
\cite{weinb} is inspiring many groups to search for edm's in a variety of
systems. (See \cite{jmphinds} for a recent review). Experiments on several
systems including the neutron \cite{glrpp}, and several species of confined
gases \cite{gas} including Radium \cite{radium}, Radon \cite{radon} and Xenon
\cite{xenon} are in various stages of preparation. These experiments are all
hoping to reach sensitivities in the range of $10^{-27}-10^{-28}e-cm$.
\ Sensitivity in this range has already been achieved in the case of $Hg$
\cite{hg}. The experiments proposed represent a broad range of operating
conditions, from room temperature gases with buffer gas to laser cooled atoms
in a MOT.

In order to achieve the target sensitivities it will be necessary to deal with
the systematic error resulting from the interaction of the well-known
$\overrightarrow{v}\times\overrightarrow{E}$ field with magnetic field
gradients. Often referred to as the geometric phase effect \cite{cumm},
\cite{jmp} this interaction produces a frequency shift linear in the electric
field, mimicking an edm. This systematic effect is highly dependent on the
operating conditions of the experiment. While experiments in small vessels and
with high pressure buffer gas are expected to be relatively insensitive to the
systematic effect, each of the proposed experiments will have to be analyzed
in detail to judge its sensitivity to the effect and to find methods of
dealing with it. In this work we introduce an analytic form of the correlation
function which determines the behavior of the frequency shift \cite{LG} and
show in detail how it depends on the operating conditions of the experiment.
For clarity we specialize the discussion to to the Los Alamos proposal for a
neutron edm search using Ultra-cold neutrons (UCN) and $He^{3}$ atoms
diffusing in superfluid $He^{4}$ as a co-magnetometer, \cite{lanledm} but the
generalization to other cases is straightforward.

First analyzed by Commins \cite{cumm} in the context of a beam experiment, the
frequency shift has been discussed in some detail by Pendlebury et al
\cite{jmp} in connection with experiments involving stored particle gases.
Additional discussion and calculations have been given by \cite{LG}.

Our present understanding of the effect can best be summarized by figure 1,
which appeared as figure 3 in \cite{LG}. This is a plot of the normalized
(linear in $E$) frequency shift vs. normalized Larmor frequency for various
values of collision mean free path and wall specularity.

These results have been obtained by numerical simulation of the
position-velocity correlation function and taking the Fourier transform.
According to [\cite{LG}, equ 26] the frequency shift is given by
\begin{equation}
\delta\omega=ab\lim_{t\rightarrow\infty}\int_{0}^{t}d\tau R\left(
\tau\right)  \cos\omega_{o}\tau\label{2}%
\end{equation}
where $R\left(  \tau\right)  $ is the position velocity correlation function
defined in [\cite{LG}, equ. 27]:%
\begin{equation}
R\left(  \tau\right)  =\left\langle \overrightarrow{r}(t)\cdot\overrightarrow
{v}(t-\tau)-\overrightarrow{r}(t-\tau)\cdot\overrightarrow{v}(t)\right\rangle
\label{1a}%
\end{equation}
and $a=\frac{\gamma}{2}\partial B_{z}/\partial z,$ $b=$ $\gamma E/c$. From the
experimental point of view it is very appealing to try to make use of the zero
crossing, apparent in figure 1, to reduce the effect.

In this note we present an analytic model for the correlation function
$R\left(  \tau\right)  $ and compare it to the results obtained previously by
numerical simulations. Using the limiting form of this model, valid for
collision mean free paths small compared to the vessel size, we calculate the
temperature dependence of the frequency shift for $^{3}He$ diffusing in
superfluid $^{4}He$. This is an experimentally favored regime as collisions
are seen to reduce the effect, as well as the slope at the zero crossing,
considerably. We also propose a method for measuring the spectrum of the
correlation function, \emph{i.e.} the frequency dependence of the shift, directly.

\section{Analytical model for the correlation function $R\left(  \tau\right)
$}

\subsection{Gas collisions}

1) We consider a particle that moves among scattering centers. If $\tau_{c}$
is the average time between collisions the velocity autocorrelation function
will have the form \cite{mcgreg}
\begin{equation}
\psi(t)\equiv\langle\vec{v}(t)\vec{v}(0)\rangle=v^{2}\,e^{-\frac{\tau}%
{\tau_{c}}}. \label{4}%
\end{equation}
In the other words, $\psi(t)$ obeys the equation:
\begin{equation}
\frac{d\psi(t)}{dt}+\frac{1}{\tau_{c}}\psi(t)=0. \label{5}%
\end{equation}

\subsection{Wall collisions}

We consider particles moving in a cylindrical storage cell in a vacuum. As
shown in \cite{jmp},\cite{LG} the frequency shift depends only on the motion
in the $x,y$ plane. Referring to figure 2, the trajectory sweeps out an angle
\[
\alpha=\arccos\left(  r/R\right)
\]
with respect to the center in a time
\[
\frac{\tau_{wall}}{2}=\frac{\sqrt{R^{2}-r^{2}}}{v}%
\]
where $\tau_{wall}$ is the time between wall collisions. The average angular
velocity for a single trajectory is then%
\[
\omega=\frac{\arccos\left(  r/R\right)  v}{\sqrt{R^{2}-r^{2}}}.
\]
The average squared frequency for all particles with velocity $v$ (in the
$x,y$ plane) has been given by \cite{jmp}, equ, (28)\ as%
\begin{equation}
\left\langle \omega_{o}^{2}\right\rangle =\frac{\pi^{2}}{6}\left(  \frac
{v^{2}}{R^{2}}\right)  \label{16}%
\end{equation}
As the motion is not strictly circular each trajectory will experience a
complicated spectrum for the time varying field (Ref. \cite{jmp}, section IV D
and figure 7), but the mean square of the fundamental frequency will be given
by (\ref{16}).

Assuming that the result is dominated by the fundamental frequency, the
velocity autocorrelation function will, to this approximation , obey the
equation:
\begin{equation}
\frac{d^{2}\psi(t)}{dt^{2}}+\left\langle \omega_{o}^{2}\right\rangle
\psi(t)=0, \label{17}%
\end{equation}

\subsubsection{Non-specular wall collisions}

Following the proceeding section we are modelling the correlation function as
a harmonic oscillator. During one traversal of the cell the oscillator will
undergo a phase change
\[
\phi=\omega\tau_{wall}=2\alpha=2\arccos\left(  r/R\right)
\]
A non-specular reflection from the wall would result in a change in the
incident angle for the next collision, $\chi$, by a random amount $\Delta\chi$
and hence a change in the accumulated oscillator phase by
\begin{align*}
\Delta\phi &  =\frac{d\phi}{d\chi}\Delta\chi\\
\frac{d\phi}{d\chi}  &  =\frac{d\phi}{dr}\frac{dr}{d\chi}%
\end{align*}
With
\begin{align*}
\sin\chi &  =r/R\\
\cos\chi &  =\sqrt{1-r^{2}/R^{2}}%
\end{align*}
we have
\begin{align*}
\frac{d\chi}{dr}  &  =\frac{1}{\sqrt{R^{2}-r^{2}}}\\
\frac{d\phi}{dr}  &  =\frac{2}{R\sqrt{1-r^{2}/R^{2}}}%
\end{align*}
so%
\[
\Delta\phi=2\Delta\chi
\]
Since the changes $\Delta\phi$ are random the phase $\phi$ will make a random
walk so that after a time $t$ we will have
\[
\left\langle \left(  \Delta\phi\right)  ^{2}\right\rangle _{t}=4\left\langle
\left(  \Delta\chi\right)  ^{2}\right\rangle \frac{t}{\tau_{wall}%
}=2\left\langle \left(  \Delta\chi\right)  ^{2}\right\rangle \frac{tv}%
{\sqrt{R^{2}-r^{2}}}%
\]
Averaging the amplitude of the oscillator over the distribution of $\Delta
\phi$ the amplitude will be reduced by
\[
\left\langle \cos\phi\right\rangle =1-\frac{1}{2}\left\langle \left(
\Delta\phi\right)  ^{2}\right\rangle _{t}\equiv1-\frac{t}{\tau_{non-spec}}%
\sim\exp\left(  -\frac{t}{\tau_{non-spec}}\right)
\]
Thus we have
\[
\frac{1}{\tau_{non-spec}}=\frac{\left\langle \left(  \Delta\chi\right)
^{2}\right\rangle v}{\sqrt{R^{2}-r^{2}}}.
\]
Averaging over $r$ we find%
\[
\left\langle \frac{1}{\tau_{non-spec}}\right\rangle =\left\langle \left(
\Delta\chi\right)  ^{2}\right\rangle v\int_{0}^{R}\frac{F\left(  r\right)
dr}{\sqrt{R^{2}-r^{2}}}=\frac{4}{\pi}\left\langle \left(  \Delta\chi\right)
^{2}\right\rangle \frac{v}{R},
\]
where $F\left(  r\right)  dr$ is the probability that a trajectory has a
pericentric distance between $r,r+dr:$%
\[
F\left(  r\right)  =\frac{4\sqrt{R^{2}-r^{2}}}{\pi R^{2}}.
\]

\subsection{Model expression for the correlation function}

When the particles are moving inside a vessel containing scatterers, one can
expect that the equation for the velocity correlation function is the
combination of (\ref{5}) and (\ref{17}), thus we are led to write
\begin{equation}
\frac{d^{2}\psi(\tau)}{d\tau^{2}}+\frac{1}{\tau_{c}}\frac{d\psi(\tau)}{d\tau
}+\left\langle \omega_{0}^{2}\right\rangle \psi(\tau)=0.
\end{equation}
This is just the equation for a damped harmonic oscillator and its general
solution is of the form:
\begin{equation}
\psi(\tau)=c_{1}e^{-\eta_{1}\tau}+c_{2}e^{-\eta_{2}\tau},
\end{equation}
where
\begin{equation}
\eta_{1}=\frac{1}{2\tau_{c}}+\sqrt{\frac{1}{4\tau_{c}^{2}}-\left\langle
\omega_{0}^{2}\right\rangle },\qquad\eta_{2}=\frac{1}{2\tau_{c}}-\sqrt
{\frac{1}{4\tau_{c}^{2}}-\left\langle \omega_{0}^{2}\right\rangle }.
\end{equation}

Then, we have for the boundary condition satisfied by $\psi(\tau),$ \cite{LG}%
\[
h(\tau)=\int\limits_{0}^{\tau}\psi(t)dt=\frac{R(\tau)}{2}\rightarrow
0,\qquad\mbox{when}\qquad\tau\rightarrow\infty.
\]

Thus, the velocity correlation function $\left[  \psi(\tau)=\langle\vec
{v}(t)\vec{v}(0)\rangle\right]  $ will have the form:
\begin{equation}
\psi(\tau)=\frac{\eta_{1}v^{2}}{\eta_{1}-\eta_{2}}\left(  e^{-\eta_{1}\tau
}-\frac{\eta_{2}}{\eta_{1}}e^{-\eta_{2}\tau}\right)  .
\end{equation}

and the function $R(\tau)$ will be given by:
\begin{equation}
R(\tau)=2\int\limits_{0}^{\tau}\psi(x)dx=\frac{2v^{2}}{\eta_{1}-\eta_{2}%
}\left(  1-e^{-(\eta_{1}-\eta_{2})\tau}\right)  e^{-\eta_{2}\tau}. \label{22}%
\end{equation}

\subsubsection{Overdamped, short mean free path, limit}

In the overdamped limit
\begin{equation}
\frac{1}{2\tau_{c}}\gg\omega_{0}%
\end{equation}
we find:
\begin{equation}
\eta_{1}\simeq\frac{1}{\tau_{c}},\qquad\eta_{2}\simeq\tau_{c}\left\langle
\omega_{0}^{2}\right\rangle ,\qquad\eta_{1}\gg\eta_{2}. \label{19}%
\end{equation}
Therefore:
\begin{equation}
R(\tau)=2\lambda v\left(  1-e^{-\frac{\tau}{\tau_{c}}}\right)  e^{-\frac{\tau
}{T}}, \label{18}%
\end{equation}
with
\begin{equation}
T=\frac{1}{\tau_{c}\left\langle \omega_{0}^{2}\right\rangle }=\left(  \frac
{6}{\pi^{2}}\right)  \frac{R^{2}}{v\lambda}\,. \label{20}%
\end{equation}
using (\ref{16}).

The numerical simulations of the correlation function shown in figure 1 of
\cite{LG} (for a cylinder), were seen to be in good agreement (in the short
mean free path limit) with equ. (\ref{18}) above (equation 43, \cite{LG})
with
\begin{equation}
T=.6R^{2}/\lambda v \label{3}%
\end{equation}
where $\lambda$ is the collision mean free path, $v$ is the particle velocity
and $R$ is the radius of the cylindrical measurement cell, (taken as $R=25$,
in the rest of this paper). The factor $0.6$ in equation (\ref{3}) is in good
agreement with the factor $\left(  6/\pi^{2}=\allowbreak0.608\right)  $ in
equation (\ref{20}).

In the case of $He^{3}$ moving in superfluid $He^{4},\tau_{c},v$ and $\lambda$
are all functions of temperature For UCN on the other hand the only collisions
are those with the walls, all parameters are independent of temperature and we
have to take the limit%
\[
\frac{1}{2\tau_{c}}\ll\omega_{0}.
\]
In figure 3 we show a comparison of the correlation functions calculated by
numerical simulation, red, the model equ. (\ref{22}), blue and the short mean
free path limit (\ref{18}), green.

In the figures $r_{o}=R/\lambda$. We see the model gives a reasonable fit for
a wide range of $r_{o}$ and the short mean free path expression (\ref{18}),
(green), is a reasonable fit for $r_{o}\gtrsim2.5$. The disagreement for
smaller $r_{o}$ is most likely due to the neglect in our model of higher
harmonics of the motional frequency

In the following we will use (\ref{18}) to calculate the frequency shift
averaged over a Maxwell-Boltzman velocity distribution as a function of
temperature for $^{3}He$ diffusing in $^{4}He$, taking into account the
velocity dependence of the mean free path.

\subsection{ Calculation of the zero crossing frequency for the v$\times E$
induced systematic\label{section24}}

According to (\ref{2}) the frequency shift is proportional to $S\left(
\omega\right)  $, the \ cosine Fourier transform of $R\left(  \tau\right)  $
which is given by (\ref{1a})$.$ For $\exp\left(  -\alpha\left|  \tau\right|
\right)  $ the cosine Fourier transform is $\alpha/\left(  \alpha^{2}%
+\omega^{2}\right)  $ so that we have, using the short mean free path limit of
the model, equation (\ref{18})%
\begin{equation}
S(\omega)=\left[  \frac{\eta_{2}}{\omega^{2}+\eta_{2}^{2}}-\frac{\eta_{1}%
+\eta_{2}}{\omega^{2}+\left(  \eta_{1}+\eta_{2}\right)  ^{2}}\right]  \lambda
v \label{6}%
\end{equation}
This is easily seen to have the correct limits for $\omega\rightarrow0,\infty
$, (equations 54 and 70 in \cite{LG}). To find the value of $\omega$ where the
effect goes to zero we put
\begin{align*}
S\left(  \omega\right)   &  =0\\
\omega^{2}+\left(  \eta_{1}+\eta_{2}\right)  ^{2}  &  =\left(  \omega^{2}%
+\eta_{2}^{2}\right)  \frac{\left(  \eta_{1}+\eta_{2}\right)  }{\eta_{2}}\\
\omega &  =\sqrt{\eta_{1}\eta_{2}+\eta_{2}^{2}}\approx\sqrt{\eta_{1}\eta_{2}}%
\end{align*}
in the case where $\eta_{1}\gg\eta_{2}\quad\left(  \lambda\ll R\right)  .$
Thus we see that the zero crossing, in the limit, is at
\[
\omega_{o}\approx\sqrt{\eta_{1}\eta_{2}}=\sqrt{\frac{\lambda v}{.6R^{2}%
\tau_{c}}}=1.3\frac{v}{R}%
\]
which should hold for reasonably small values of $\lambda/R$. Thus for short
$\lambda$ all curves should have the same zero crossing. This is seen to be
satisfied by the numerical simulations of figure 1.

\section{Frequency shift averaged over velocity distribution}

Using the limiting analytical model (\ref{18}) for the correlation function
and its Fourier transform we can perform an average over the velocity
distribution. We do this for the case (1) of a constant collision man free
path and then for the realistic case (2) of a mean free path proportional to
velocity in order to show the influence of a variable collision mean free path.

\subsection{Mean free path independent of velocity}

The frequency shift, is given in terms of the spectrum $S(\omega)$ (\ref{6})
by:%
\[
\delta\omega(\omega,T)=abS(\omega)
\]%

\begin{equation}
\delta\omega(\omega,T)=abR^{2}\left(  .6\right)  \left[  \frac{\left(
k(T)/.6\right)  ^{2}}{\left(  \omega R/v\right)  ^{2}+\left(  k(T)/.6\right)
^{2}}-\frac{\left(  k(T)/.6\right)  \left(  k(T)/.6+1/k(T)\right)  }{\left(
\omega R/v\right)  ^{2}+\left(  k(T)/.6+1/k(T)\right)  ^{2}}\right]
\label{00}%
\end{equation}
with $k=\lambda/R=1/r_{o}$ being a function of temperature $\left(
\lambda(T)=3D(T)/v(T)\right)  $ as shown in figure 3c), ($D(T)=1.6/T^{7}$ is
the diffusion coefficient for $He^{3}$ in $He^{4}$ as measured by
\cite{lametal}, and we have taken $R=25$) .

The average of this over the velocity distribution of the $He^{3}$ will be
given by%
\begin{align*}
\left\langle \delta\omega(\omega,T)\right\rangle  &  =\frac{4}{\sqrt{\pi}}%
\int\delta\omega(\omega,T)\frac{v^{2}}{\beta^{2}}e^{-v^{2}/\beta^{2}}\frac
{dv}{\beta}\\
&  =abR^{2}\left(  .6\right)  \frac{4}{\sqrt{\pi}}\int_{0}^{\infty}\left[
\frac{1}{\left(  \omega R/\alpha_{2}\right)  ^{2}+v^{2}}-\frac{\alpha
_{2}/\alpha_{1}}{\left(  \omega R/\alpha_{1}\right)  ^{2}+v^{2}}\right]
v^{4}e^{-v^{2}/\beta^{2}}\frac{dv}{\beta^{3}}%
\end{align*}
where $\alpha_{2}=\left(  k(T)/.6\right)  $, $\alpha_{1}=\left(
k(T)/.6+1/k(T)\right)  .$ Now we have%
\begin{equation}
\frac{4}{\sqrt{\pi}}\int_{0}^{\infty}\frac{x^{4}}{a^{2}+x^{2}}e^{-x^{2}%
/\beta^{2}}dx\allowbreak\allowbreak\equiv F(a,\beta)=\allowbreak2a^{3}%
\sqrt{\pi}e^{\frac{a^{2}}{\beta^{2}}}\left(  1-\operatorname{erf}\left(
\frac{a}{\beta}\right)  \right)  -2a^{2}\allowbreak\beta+\beta^{3} \label{21}%
\end{equation}
so that%
\begin{align*}
\left\langle \delta\omega(\omega,T)\right\rangle  &  =\frac{abR^{2}}{\beta
^{3}}\left(  .6\right)  \left[  F\left(  \frac{\omega R}{\alpha_{2}}%
,\beta(T)\right)  -\frac{\alpha_{2}}{\alpha_{1}}F\left(  \frac{\omega
R}{\alpha_{1}},\beta(T)\right)  \right] \\
&  \equiv abR^{2}\Omega\left(  \omega,T\right)
\end{align*}
where $\beta(T)=1.28\times10^{4}\sqrt{\frac{T}{7.2}}cm/\sec$ is the most
probable velocity for $He^{3}$ in $He^{4}$. $F(a,\beta)$ tends to be difficult
to calculate numerically for $\alpha/\beta\gtrsim5$ so we will use the
asymptotic expansion (Ref. \cite{abramsteg}, section 7.1.23, pg 298)%
\[
\sqrt{\pi}e^{z^{2}}\operatorname{erf}c(z)\rightarrow\left(  \frac{1}%
{z}\right)  \left(  1-\frac{1}{2z^{2}}+\frac{3}{4z^{4}}-\frac{15}{8z^{6}%
}+....\right)
\]
for $\alpha/\beta>4.$ Note that the leading term in the expansion is canceled
by the term $2a^{2}\allowbreak\beta$ in (\ref{21}). To plot the results we
choose
\begin{equation}
X\left(  \omega,T\right)  =\omega R/\beta(T) \label{7}%
\end{equation}
as the independent variable.

Figure 4 shows the normalized, averaged over the velocity distribution,
frequency shift plotted versus the frequency relative to the appropriate, most
probable velocity, (\ref{7}):

\bigskip Expanding the plot we see the region near the zero crossing (figure 5).

and expanding still further (figure 6)

In this figure we see the zero crossings converging as the mean free path gets
smaller (increasing temperature) just as in the single velocity case (section
\ref{section24}).

From the point of view of an experiment it is perhaps more interesting to plot
the results as a function of temperature.

Figure 7 shows the temperature dependence of the velocity averaged frequency
shift for a range of frequencies specified by the corresponding values of
$X\left(  \omega,T\right)  $, (equ. \ref{7}), i.e. $\omega=\frac{\beta(T)}%
{R}X\left(  \omega,T\right)  $.

Expanding the plot (figure 8) shows the region where the effect can be minimized:

Here we see that the applied frequency can be chosen so that the velocity
averaging does indeed reduce the effect. In fact reductions of $10^{-5}$ to
$10^{-6}$ appear feasible.

\subsection{Mean free path $\propto velocity,$ cross section $\sim1/v$}

We have $k(T)=\lambda(T)/R.$ \ Since the velocity of the $He^{3}$ is much less
than the phonon velocity ($2.2\times10^{4}cm/\sec$) the collision rate of
phonons with the $He^{3}$ will be independent of the $He^{3}$ velocity. In a
time $\tau_{c}$ a $He^{3}$ with velocity $v$, will move a distance
$\lambda_{v}(T)=v\tau_{c}(T).$ Thus
\[
k_{v}(T)=\lambda_{v}(T)/R=v\tau_{c}(T)/R=v/s(T)
\]
with $s(T)=R/\tau_{c}(T)$.

From equation (\ref{00})
\begin{equation}
\delta\omega(\omega,T)=abR^{2}\left(  .6\right)  \left[  \frac{\left(
k_{v}(T)/.6\right)  ^{2}}{\left(  \omega R/v\right)  ^{2}+\left(
k_{v}(T)/.6\right)  ^{2}}-\frac{\left(  k_{v}(T)/.6\right)  \left(
k_{v}(T)/.6+1/k_{v}(T)\right)  }{\left(  \omega R/v\right)  ^{2}+\left(
k_{v}(T)/.6+1/k_{v}(T)\right)  ^{2}}\right]
\end{equation}
The average of this over the velocity distribution of the $He^{3}$ will be
given by%
\begin{align}
\left\langle \delta\omega(\omega,T)\right\rangle  &  =\frac{4}{\sqrt{\pi}}%
\int\delta\omega(\omega,T)\frac{v^{2}}{\beta^{2}}e^{-v^{2}/\beta^{2}}\frac
{dv}{\beta}\nonumber\\
&  =abR^{2}\left(  .6\right)  \frac{4}{\sqrt{\pi}}\int_{0}^{\infty}\left[
\frac{1}{\left(  \omega R/\alpha_{2}\right)  ^{2}+v^{2}}-\frac{\alpha
_{2}/\alpha_{1}}{\left(  \omega R/\alpha_{1}\right)  ^{2}+v^{2}}\right]
v^{4}e^{-v^{2}/\beta^{2}}\frac{dv}{\beta^{3}}\label{8}\\
&  \equiv abR^{2}\Psi\left(  \omega,T\right)
\end{align}
where $\alpha_{2}=\left(  k_{v}(T)/.6\right)  $, $\alpha_{1}=\left(
k_{v}(T)/.6+1/k_{v}(T)\right)  .$ In the region of interest for suppression of
the effect, $k_{v}(T)\ll1$, we have $\alpha_{1}\sim1/k_{v}(T).$ Then the
integral in (\ref{8}) can be written
\[
=\int_{0}^{\infty}\left[  \frac{y^{2}}{\left(  \frac{.6s}{\beta(T)}\right)
^{2}\left(  \frac{\omega R}{\beta(T)}\right)  ^{2}+y^{4}}-\frac{1/.6}{\left(
\frac{\omega R}{\beta(T)}\right)  ^{2}+\frac{s(T)^{2}}{\beta(T)^{2}}}\right]
y^{4}e^{-y^{2}}dy
\]

where $y=v/\beta(T).$ The second integral can be evaluated as
\[
\int_{0}^{\infty}y^{4}e^{-y^{2}}dy=\frac{3}{2}\frac{\sqrt{\pi}}{4}%
\]

Defining
\[
x(\omega,T)=\frac{\omega R}{\beta(T)}%
\]
and writing
\[
\frac{s(T)}{\beta(T)}=\frac{R}{\beta(T)\tau_{c}(T)}=\frac{R}{\lambda_{c}%
(T)}=\frac{1}{k(T)}%
\]
where $\lambda_{c}(T)$ and $k(T)$ are evaluated at the most probable velocity
$\beta(T),$ we have $a(T)=\left(  \frac{.6s}{\beta(T)}\right)  \left(
\frac{\omega R}{\beta(T)}\right)  =\frac{.6}{k(T)}x(\omega,T)$

\bigskip

The first integral can be evaluated in terms of a hypergeometric function
($\operatorname{hypergeom}\left(  \left[  1\right]  ,\left[  \frac{1}%
{4},-\frac{1}{4}\right]  ,-\frac{1}{4}x^{2}\right)  )$ but the series for this
has some convergence difficulties for the parameters of interest. Thus we
define
\[
f(a)=\int_{0}^{\infty}\frac{y^{6}}{a^{2}+y^{4}}e^{-y^{2}}dy
\]
(to be evaluated numerically as the integrand is well behaved) so that%

\begin{equation}
\Psi\left(  \omega,T\right)  =\left(  .6\right)  \left[  \frac{4}{\sqrt{\pi}%
}f\left[  \frac{.6}{k(T)}x(\omega,T)\right]  -\frac{2.5}{\left(
x(\omega,T)\right)  ^{2}+\left(  \frac{1}{k(T)}\right)  ^{2}}\right]
\label{9}%
\end{equation}
In \ figure 9 we plot the frequency shift averaged over velocity as a function
of Larmor frequency for fixed temperature taking into account the velocity
dependence of the mean free path.

Figures 10 and 11 show the frequency shift as a function of temperature for
various frequencies

\section{Measurement of the correlation function, $R\left(  \tau\right)  $}

While the theory presented here is expected to be accurate for the case of
polarized $He^{3}$ diffusing in superfluid $He^{4}$ it would be nice to be
able to confirm it experimentally and to check its applicability to other edm
searches in progress. In this section we present a generally applicable,
straightforward method to measure the frequency spectrum of the relevant
correlation function (\ref{1a}) and hence the frequency dependence of the
frequency shift directly.

If we apply a uniform magnetic field gradient $\partial B_{z}/\partial z$
large enough so that it dominates all other field gradients that may be
present, there will be a radial field in the $x-y$ plane $\overrightarrow
{B}_{\overrightarrow{r}}=-\left(  \overrightarrow{r}/2\right)  \left(
\partial B_{z}/\partial z\right)  .$ Then
\[
\frac{\partial B_{x,y}}{\partial x,y}=-\frac{1}{2}\frac{\partial B_{z}%
}{\partial z}=-a
\]
and the field correlation functions will be%
\[
\left\langle B_{x_{i}}\left(  t\right)  B_{x_{i}}\left(  t+\tau\right)
\right\rangle =\left(  a\right)  ^{2}\left\langle x_{i}\left(  t\right)
x_{i}\left(  t+\tau\right)  \right\rangle
\]
where $x_{i}=x$ or $y$. Then following McGregor \cite{mcgreg} the relaxation
time, $T_{1}$ will be given by (in the case when $\partial B_{z}/\partial z$
is large enough so that wall relaxation can be neglected)%
\[
\frac{1}{T_{1}}=\frac{\gamma^{2}a^{2}}{2}\left[  S_{r}\left(  \omega
_{o}\right)  \right]
\]
with%
\begin{align*}
S_{r}\left(  \omega\right)   &  =\int_{-\infty}^{\infty}\left\langle
\overrightarrow{r_{\bot}}\left(  t\right)  \cdot\overrightarrow{r_{\bot}%
}\left(  t+\tau\right)  \right\rangle e^{-i\omega\tau}d\tau\\
&  =\int_{-\infty}^{\infty}R_{\overrightarrow{r}\overrightarrow{r}}\left(
\tau\right)  \cos\omega\tau d\tau
\end{align*}
where we recognize that the correlation function $R_{\overrightarrow
{r}\overrightarrow{r}}\left(  \tau\right)  $ is an even function of $\tau$. A
measurement of $T_{1}$ will thus yield the function $S_{r}\left(
\omega\right)  $. Now%
\begin{align*}
\omega^{2}S_{r}\left(  \omega\right)   &  =-\int_{-\infty}^{\infty
}R_{\overrightarrow{r}\overrightarrow{r}}\left(  \tau\right)  \frac{d^{2}%
}{d\tau^{2}}\left(  \cos\omega\tau\right)  d\tau\\
&  =-\int_{-\infty}^{\infty}\frac{d^{2}R_{\overrightarrow{r}\overrightarrow
{r}}\left(  \tau\right)  }{d\tau^{2}}\left(  \cos\omega\tau\right)  d\tau\\
&  =\int_{-\infty}^{\infty}R_{\overrightarrow{v}\overrightarrow{v}}\left(
\tau\right)  \left(  \cos\omega\tau\right)  d\tau
\end{align*}
where we used $R_{\overrightarrow{v}\overrightarrow{v}}\left(  \tau\right)
=-d^{2}R_{\overrightarrow{r}\overrightarrow{r}}\left(  \tau\right)  /d\tau
^{2}$ \cite{papoul} for the velocity correlation function. Then
\[
R_{\overrightarrow{v}\overrightarrow{v}}\left(  \tau\right)  =\frac{1}{2\pi
}\int_{-\infty}^{\infty}\omega^{2}S_{r}\left(  \omega\right)  \cos\omega\tau
d\tau
\]
By comparison with equ (38) of \cite{LG} we identify $\omega^{2}S_{r}\left(
\omega\right)  /2\pi$ with $\psi\left(  \omega\right)  $ of that work and we
then see that the systematic frequency shift is given by (equation 40 of
\cite{LG})%
\begin{equation}
\delta\omega=-\frac{ab}{2\pi}\int_{-\infty}^{\infty}\frac{\omega^{2}%
S_{r}\left(  \omega\right)  }{\left(  \omega_{o}^{2}-\omega^{2}\right)
}d\omega\label{10}%
\end{equation}
Since we can determine $S_{r}\left(  \omega\right)  $ from the $T_{1}$
measurements we can determine the frequency dependence of the systematic
frequency shift for any experimental conditions without the need of applying
an electric field.

\section{Arbitrary Magnetic field geometry}

Our discussion has assumed a magnetic field configuration with $G_{z}=\partial
B_{z}/\partial z$ constant. Pendlebury et al \cite{jmp} have shown, using a
geometric phase argument, that regardless of the field geometry the effect
only depends on the volume average of $G_{z}$ in the high frequency (called by
them the adiabatic) limit. In a recent note, Harris and Pendlebury
\cite{harrjmp} have shown that in the case of a field produced by a dipole
external to the measurement cell, this does not hold in the low frequency
(diffusion) limit. In this section we discuss this problem using our
correlation function approach in order to give some physical insight into what
is happening and display details of the transition from one case to another.

\subsection{Short time (high frequency, adiabatic) limit of the correlation function}

Reference \cite{LG}, has shown that the systematic edm is given, in general,
as the Fourier transform of a certain correlation function of the time varying
field seen by the neutrons as they move through the apparatus. Equation (23)
of that paper gives the frequency shift proportional to $E$ as $\left(
\overrightarrow{\omega}\left(  t\right)  \text{ lies in the }x,y\text{
plane}\right)  $%

\begin{equation}
\delta\omega_{E}(t)=-\frac{1}{2}\int_{0}^{t}d\tau\left\{
\begin{array}
[c]{c}%
\cos\omega_{o}\tau\left[  \overrightarrow{\omega}\left(  t\right)
\times\overrightarrow{\omega}\left(  t-\tau\right)  \right] \\
+\sin\omega_{o}\tau\left[  \omega_{x}\left(  t\right)  \omega_{x}\left(
t-\tau\right)  +\omega_{y}\left(  t-\tau\right)  \omega_{y}\left(  t\right)
\right]
\end{array}
\right\}  \label{14}%
\end{equation}

It can be shown that the term multiplying $\sin\omega_{o}\tau$ goes to zero on
averaging over a uniform velocity distribution $\left(  \left\langle
v_{x}v_{y}\right\rangle =0,\quad v_{x}^{2}=v_{y}^{2}=v^{2}/2\right)  $ and
using $\overrightarrow{\nabla}\times\overrightarrow{B}=0.$Then, for short
times, $\tau,$%

\begin{align}
\delta\omega(t)  &  =-\frac{1}{2}\int_{0}^{t}d\tau\left\{  \cos\omega_{o}%
\tau\left[  \overrightarrow{\omega}\left(  t\right)  \times\left(
\overrightarrow{\omega}\left(  t\right)  -\frac{d\overrightarrow{\omega}}%
{dt}\tau+\frac{1}{2}\frac{d^{2}\overrightarrow{\omega}}{d\tau^{2}}\tau
^{2}+...\right)  \right]  \right\} \nonumber\\
&  =-\frac{1}{2}\int_{0}^{t}d\tau\left\{  \cos\omega_{o}\tau\left[
-\overrightarrow{\omega}\left(  t\right)  \times\left(  \frac{d\overrightarrow
{\omega}}{dt}\tau-\frac{1}{2}\frac{d^{2}\overrightarrow{\omega}}{d\tau^{2}%
}\tau^{2}+...\right)  \right]  \right\}  \label{15}%
\end{align}

We are considering values of $\tau$ so small that the velocity doesn't change
in that time interval $\left(  \tau<\tau_{coll}\right)  .$

Then
\begin{align*}
\overrightarrow{\omega}(t)  &  =\gamma\left(  \overrightarrow{B}%
_{xy}(t)+\overrightarrow{v}/c\times\overrightarrow{E}\right) \\
\frac{d\overrightarrow{\omega}}{dt}  &  =\gamma\left(  \overleftrightarrow
{\triangledown B}\left(  \overrightarrow{x}\left(  t\right)  \right)
\cdot\overrightarrow{v}\right) \\
\frac{d^{2}\overrightarrow{\omega}}{d\tau^{2}}  &  =\gamma\sum_{i,j}%
\frac{\partial^{2}\overrightarrow{B}}{\partial x_{i}\partial x_{j}}v_{i}v_{j}%
\end{align*}

and
\begin{equation}
\delta\omega(t)=-\frac{\gamma}{2c}\int_{0}^{t}d\tau\cos\omega_{o}\tau\left[
-\left(  \overrightarrow{B}_{xy}(t)+\overrightarrow{v}\times\overrightarrow
{E}\right)  \times\left(  \frac{\partial\overrightarrow{\omega}}{\partial
t}\tau-\frac{1}{2}\frac{\partial^{2}\overrightarrow{\omega}}{\partial\tau^{2}%
}\tau^{2}+...\right)  \right]  \label{12}%
\end{equation}
The term linear in $\overrightarrow{E\text{ }}$ and $\tau$ is then%
\begin{align}
\delta\omega(t)  &  =-\frac{\gamma^{2}}{2c}\int_{0}^{t}d\tau\cos\omega_{o}%
\tau\left[  \left(  \overleftrightarrow{\triangledown B}\cdot\overrightarrow
{v}\right)  \tau\times\left(  \overrightarrow{v}\times\overrightarrow
{E}\right)  \right] \nonumber\\
&  \equiv\frac{\gamma E}{2}\int_{0}^{t}d\tau\cos\omega_{o}\tau\left(
\alpha\tau\right)  \label{13}%
\end{align}
defining
\[
\alpha=\frac{\gamma}{c}\left(  \overleftrightarrow{\triangledown B}%
\cdot\overrightarrow{v}\right)  \cdot\overrightarrow{v}%
\]
We have now calculated the correlation function for short times. It starts at
zero at $\tau=0$ and rises as $\alpha\tau$. Eventually it will reach a
maximum. By concentrating on the high frequency $\left(  \omega_{o}\right)  $
behavior of $\delta\omega$ the result will be independent of the details of
the maximum, depending only on $\alpha.$ Thus we can replace $\alpha\tau$ in
(\ref{13}) by $\sin\alpha\tau$ or any function with the same initial slope.
Thus we are led to take
\begin{align*}
\delta\omega(t)  &  \equiv\frac{\gamma E}{2}\lim_{\omega_{o}\rightarrow\infty
}\int_{0}^{t}d\tau\cos\omega_{o}\tau\sin\alpha\tau\\
&  =\lim_{\omega_{o}\rightarrow\infty}\frac{\gamma E}{2}\frac{\alpha}%
{\omega_{o}^{2}-\alpha^{2}}=\frac{\gamma E}{2}\frac{\alpha}{\omega_{o}^{2}}\\
&  =\frac{E}{2cB_{o}^{2}}\left(  \overleftrightarrow{\triangledown B}%
\cdot\overrightarrow{v}\right)  \cdot\overrightarrow{v}%
\end{align*}
Introducing components taking averages and using $\overrightarrow
{\triangledown}\cdot\overrightarrow{B}=0$ this reduces to%

\begin{equation}
\overline{\delta\omega}_{geo}=-Ev^{2}\frac{1}{4cB_{o}^{2}}\left\langle
\frac{\partial B_{z}}{\partial z}\right\rangle
\end{equation}
in agreement with equ. $\left(  2\right)  $ of \cite{LG} if, in that equation,
$R^{2}\omega_{r}^{2}$ is replaced by $\left\langle v^{2}\right\rangle
=v^{2}/2$. We have shown that in the adiabatic (short time) limit the
systematic (false) edm effect depends only on $\left\langle \frac{\partial
B_{z}}{\partial z}\right\rangle $ regardless of the geometry of the magnetic
field, a result obtained previously by Pendlebury et al \cite{jmp} and
confirmed in \cite{harrjmp}.

The next order term in (\ref{12}) is easily seen to be of order $v^{3}\tau
^{2}$ and so will average to zero, the next term which contributes will be of
order $v^{4}\tau^{3}$ and so will be negligible in the short time limit we are
considering. The condition for this to be valid is $\left(  v\tau
/\Lambda\right)  ^{2}\ll1$ where $\Lambda$ is the scale of variations in the
applied magnetic field $\left(  \frac{\partial B_{z}}{\partial z}\frac
{1}{B_{z}}\sim\Lambda^{-1}\right)  $

\subsection{Longer time behavior of the correlation function}

For long times the expansion (\ref{15}) is clearly not valid and we must
expand in a series in the spatial coordinates. We start from%
\[
\delta\omega=-\frac{\gamma^{2}}{2}\int d\tau\cos\omega_{o}\tau\left\langle
\overrightarrow{B}^{\prime}\left(  t\right)  \times\overrightarrow{B}^{\prime
}\left(  t-\tau\right)  \right\rangle _{z}%
\]
where $b=E/c$, the brackets represent an ensemble average and
\begin{align*}
B_{x}^{\prime}  &  =B_{x}\left(  \overrightarrow{r}\left(  t\right)  \right)
-bv_{y}\\
B_{y}^{\prime}  &  =B_{y}\left(  \overrightarrow{r}\left(  t\right)  \right)
+bv_{x}%
\end{align*}
Then we write%
\begin{equation}
\left\langle \overrightarrow{B}^{\prime}\left(  t\right)  \times
\overrightarrow{B}^{\prime}\left(  t-\tau\right)  \right\rangle _{z}%
=b\left\langle
\begin{array}
[c]{c}%
B_{x}\left(  \overrightarrow{r}\left(  t\right)  \right)  v_{x}\left(
t-\tau\right)  -v_{y}\left(  t\right)  B_{y}\left(  \overrightarrow{r}\left(
t-\tau\right)  \right) \\
-\left(  B_{x}\left(  \overrightarrow{r}\left(  t-\tau\right)  \right)
v_{x}\left(  t\right)  -v_{y}\left(  t-\tau\right)  B_{y}\left(
\overrightarrow{r}\left(  t\right)  \right)  \right)
\end{array}
\right\rangle \label{11}%
\end{equation}
and expand the field in a Taylor series%

\begin{align*}
B_{x}\left(  \overrightarrow{r}\left(  t\right)  \right)  =  &  \left(
B_{x}\left(  0,0,0,t\right)  +\left.  \frac{\partial B_{x}}{\partial
x}\right|  _{o}x\left(  t\right)  +\left.  \frac{\partial B_{x}}{\partial
y}\right|  _{o}y\left(  t\right)  +\left.  \frac{\partial B_{x}}{\partial
z}\right|  _{o}z\left(  t\right)  \right)  +\\
&  +\left(  \left.  \frac{\partial^{2}B_{x}}{\partial x^{2}}\right|  _{o}%
x^{2}\left(  t\right)  +\left.  \frac{\partial^{2}B_{x}}{\partial y^{2}%
}\right|  _{o}y^{2}\left(  t\right)  +\left.  \frac{\partial^{2}B_{x}%
}{\partial z^{2}}\right|  _{o}z^{2}\left(  t\right)  \right) \\
&  +\left(  \left.  \frac{\partial^{2}B_{x}}{\partial x\partial y}\right|
_{o}y(t)x\left(  t\right)  +\left.  \frac{\partial^{2}B_{x}}{\partial
y\partial z}\right|  _{o}z\left(  t\right)  y\left(  t\right)  +\left.
\frac{\partial^{2}B_{x}}{\partial z\partial x}\right|  _{o}x\left(  t\right)
z\left(  t\right)  \right) \\
&  +\left(  \left.  \frac{\partial^{3}B_{x}}{\partial x^{3}}\right|  _{o}%
x^{3}\left(  t\right)  +\left.  \frac{\partial^{3}B_{x}}{\partial y^{3}%
}\right|  _{o}y^{3}\left(  t\right)  +....\right)
\end{align*}
(similarly for $B_{y}$). Concentrating on the first and last terms in
(\ref{11}) and noting that there are no correlations between any functions
$f(x_{i},v_{i})$ and $g\left(  x_{j},v_{j}\right)  $ we find%
\[
\sum_{x_{i}=x,y}\left\langle \left(  \left.  \frac{\partial B_{x_{i}}%
}{\partial x_{i}}\right|  _{o}x_{i}\left(  t\right)  +\left.  \frac
{\partial^{2}B_{x_{i}}}{\partial x_{i}^{2}}\right|  _{o}x_{i}^{2}\left(
t\right)  +\left.  \frac{\partial^{3}B_{x}}{\partial x_{i}^{3}}\right|
_{o}x_{i}^{3}\left(  t\right)  ...\right)  v_{x_{i}}\left(  t-\tau\right)
-\left\{  \left(  t\right)  \Leftrightarrow\left(  t-\tau\right)  \right\}
\right\rangle
\]
where the second term is obtained from the first by interchanging $\left(
t\right)  $ and $\left(  t-\tau\right)  $.

By symmetry we see that $\left\langle x_{i}^{2}\left(  t\right)  v_{x_{i}%
}\left(  t-\tau\right)  \right\rangle =0$ so that the next contributing term
will be proportional to
\[
\left.  \frac{\partial^{3}B_{x}}{\partial x_{i}^{3}}\right|  _{o}\left\langle
x_{i}^{3}\left(  t\right)  v_{x_{i}}\left(  t-\tau\right)  \right\rangle
\]
The first order term will be proportional to
\[
\frac{\partial B_{x}}{\partial x}+\frac{\partial B_{y}}{\partial y}%
=-\frac{\partial B_{z}}{\partial z}%
\]
We see that the condition
\[
\left.  \frac{\partial^{3}B_{x_{i}}}{\partial x_{i}^{3}}\right|  _{o}R^{2}%
\ll\left.  \frac{\partial B_{x_{i}}}{\partial x_{i}}\right|  _{o}%
\quad\emph{or\quad}\frac{R^{2}}{\Lambda^{2}}\ll1
\]
will insure that the higher order terms can be neglected. In the extreme case
considered by Harris and Pendlbury, \cite{harrjmp}, this condition is strongly
violated so our method cannot be applied since the higher order terms remain significant.

\section{Discussion}

We see that in both the case of a velocity independent mean free path and one
proportional to velocity the behavior of the systematic frequency shift is
essentially unchanged. What is changed is the frequency where the effect goes
to zero, i.e reduced frequency $\omega^{\prime}\simeq1.6$ in the case (1) of a
constant mean free path and $\omega^{\prime}\simeq2$ for the case (2) of a
mean free path proportional to velocity. This is because longer mean free
paths contribute more to the frequency shift as seen from figure 1, and thus
higher velocities contribute more to case (2). This shifts the effective
frequency to higher values for the same value of reduced frequency.

Due to the heavy mass and slow velocity of the $He^{3}$, Baym and Ebner
\cite{baymebn} conclude that the phonon scattering on $He^{3}$ is
predominantly elastic. Single phonon absorption is kinematically forbidden on
a single $He^{3}$ and can only take place as a result of $He^{3}-He^{3}$
collisions which will be negligible for the low $He^{3}$ densities considered
here. Thus our approach, where we calculate the correlation function for an
ensemble of trajectories with constant $He^{3}$ velocity and then average the
result over the velocity distribution should be an excellent approximation.

Note that for temperatures $T\gtrsim.38K,\quad r_{o}=1/k\gtrsim20,$ so that
according to fig.3 the limiting expression for the correlation function should
be quite accurate and figures 10 and 11, which are based on an exact average
over the Maxwell distribution of the $He^{3}$ velocities, imply that one
should be able to control the effect to high degree.

What would be ideal, would be to set the size of the shift for $He^{3}$ to be
slightly negative \emph{i.e.} equal in sign and magnitude with the shift for
UCN. The UCN, with their relatively low value of $\omega_{r}$ ($\omega
_{o}/\omega_{r}\gg1),$ have a frequency shift which is represented by a point
far to the right in figure 1).

The limits on the ability to do this will be the usual experimental ones of
the temperature and magnetic field gradient stability.

We have shown that by varying the gradients to larger values it will be
possible to measure the spectrum of the velocity correlation function
directly, thus allowing a precise determination of the frequency dependence of
the frequency shift under the exact experimental conditions.

\bigskip

\section{Figure Captions}

Figure 1. Note the curves are for a single fixed velocity. The velocity
dependence is contained in the normalization of the frequency scale,
$\omega_{r}=v/R$.

\bigskip

Figure 2. Trajectory of a particle confined in a cylindrical vessel.

\bigskip

Figure 3a. Comparison of correlation function, $R\left(  \tau\right)  $ from
numerical simulations and from the analytic model. red - numerical simulation,
blue - complete model equation (\ref{22}), green - short mean free path limit
equation (\ref{18}) where appropriate. $r_{o}=R/\lambda=50,10.$

\bigskip

Figure 3b) As figure 3a) $r_{o}=0.5,1,2.5,6,7.5$

Figure 3c. Dependence of $k=\lambda(T)/R$ on temperature.

\bigskip

Figure 4. Normalized, velocity averaged fequency shift vs. normalized
frequency: $X\left(  \omega,T\right)  =\omega R/\beta\left(  T\right)  $. red
T=0.3K, blue 0.35K, green 0.4K, purple 0,45K

\bigskip

Figure 5. Normalized, velocity averaged fequency shift vs. normalized
frequency, expanded scale. red T=0.3K, blue 0.35K, green 0.4K, purple 0,45K

\bigskip

Figure 6. Normalized, velocity averaged fequency shift vs. normalized
frequency. Same as figure 8, but expanded further. red T=0.3K, blue 0.35K,
green 0.4K, purple 0,45K

\bigskip

Figure 7. Velocity averaged frequency shift as a function of temperature, K,
for various frequencies., red X=1.4, blue X=1.5, green X=1.7, purple X=1.8

\bigskip

Figure 8. Velocity averaged frequency shift as a function of temperature, K,
for various frequencies. As figure 7, scale expanded to show experimetnally
interesting region, red X=1.4, blue X=1.5, green X=1.6, purple X=1.7, light
blue X=1.8

\bigskip

Figure 9. Frequency shift averaged over velocity distribution, allowing for
velocity dependence of the $He^{3}$ mean free path, for various temperatures.
$X$ is the normalized frequency, $X=\omega R/\beta(T)$ with $\beta(T)$ the
most probable $He^{3}$ velocity. green T=0.35K, purple T=0.38K, red T=0.4K,
blue T=0.43K.

\bigskip

Figure 10. Frequency shift averaged over velocity distribution, allowing for
velocity dependence of the $He^{3}$ mean free path, versus temperature, K, for
various frequencies specified by the value of $X=\omega R/\beta$ with $\beta$
the most probable $He^{3}$ velocity. blue X=3, green X=2.2, red X=2, purple X=1.8.

\bigskip

Figure 11. Expanded plot of the frequency shift averaged over velocity
distribution, allowing for velocity dependence of the $He^{3}$ mean free path,
versus temperature, K, for various frequencies specified by the value of
$X=\omega R/\beta(T)$ with $\beta(T)$ the most probable $He^{3}$ velocity.
blue X=3, green X=2.2, red X=2, purple X=1.8.
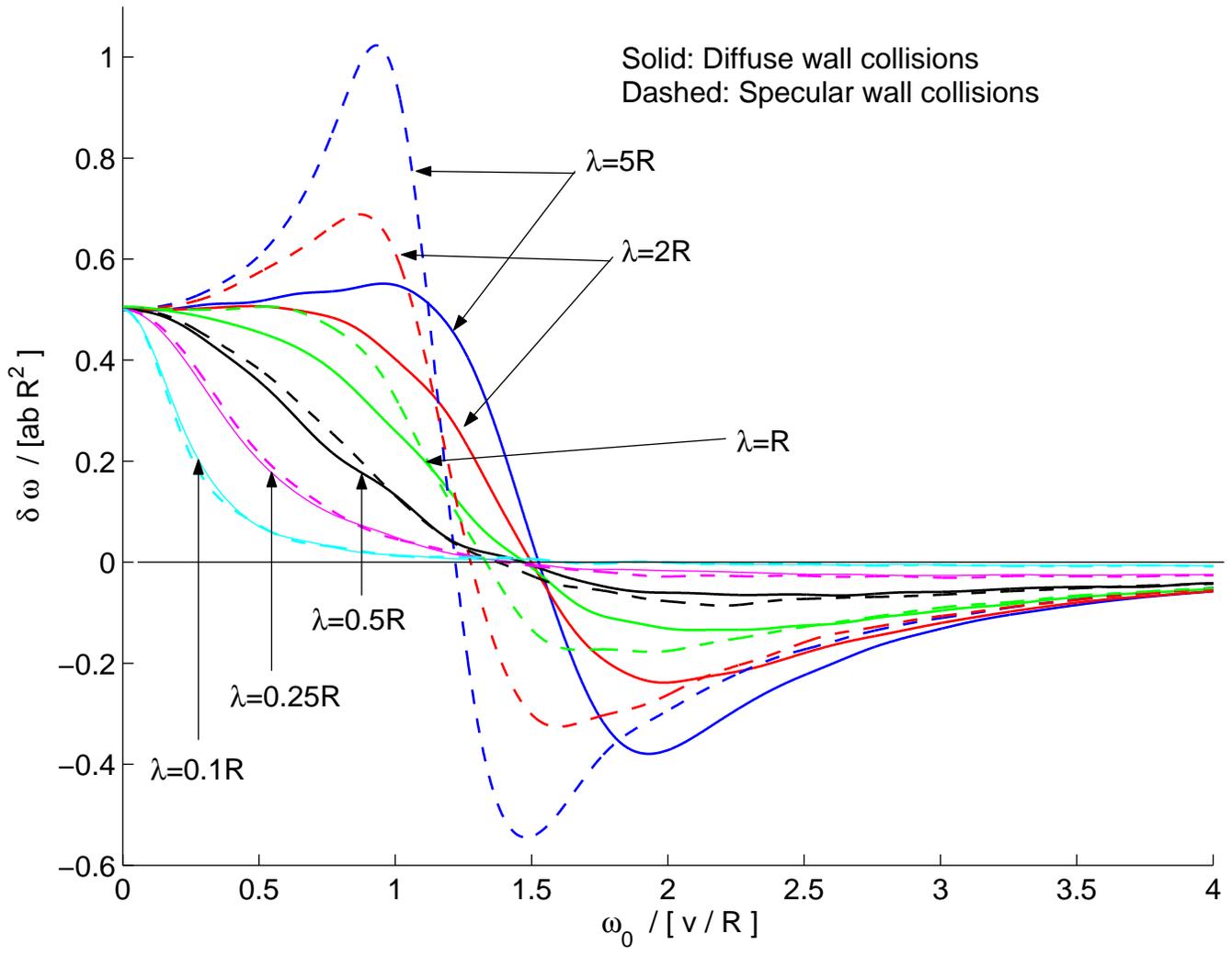
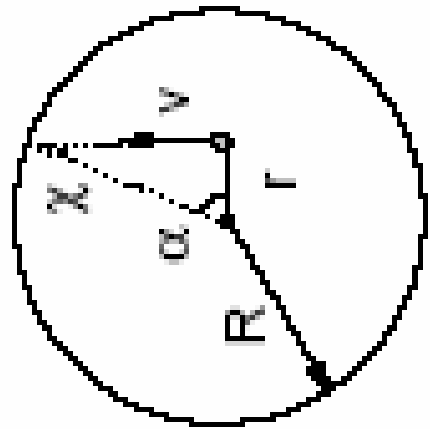
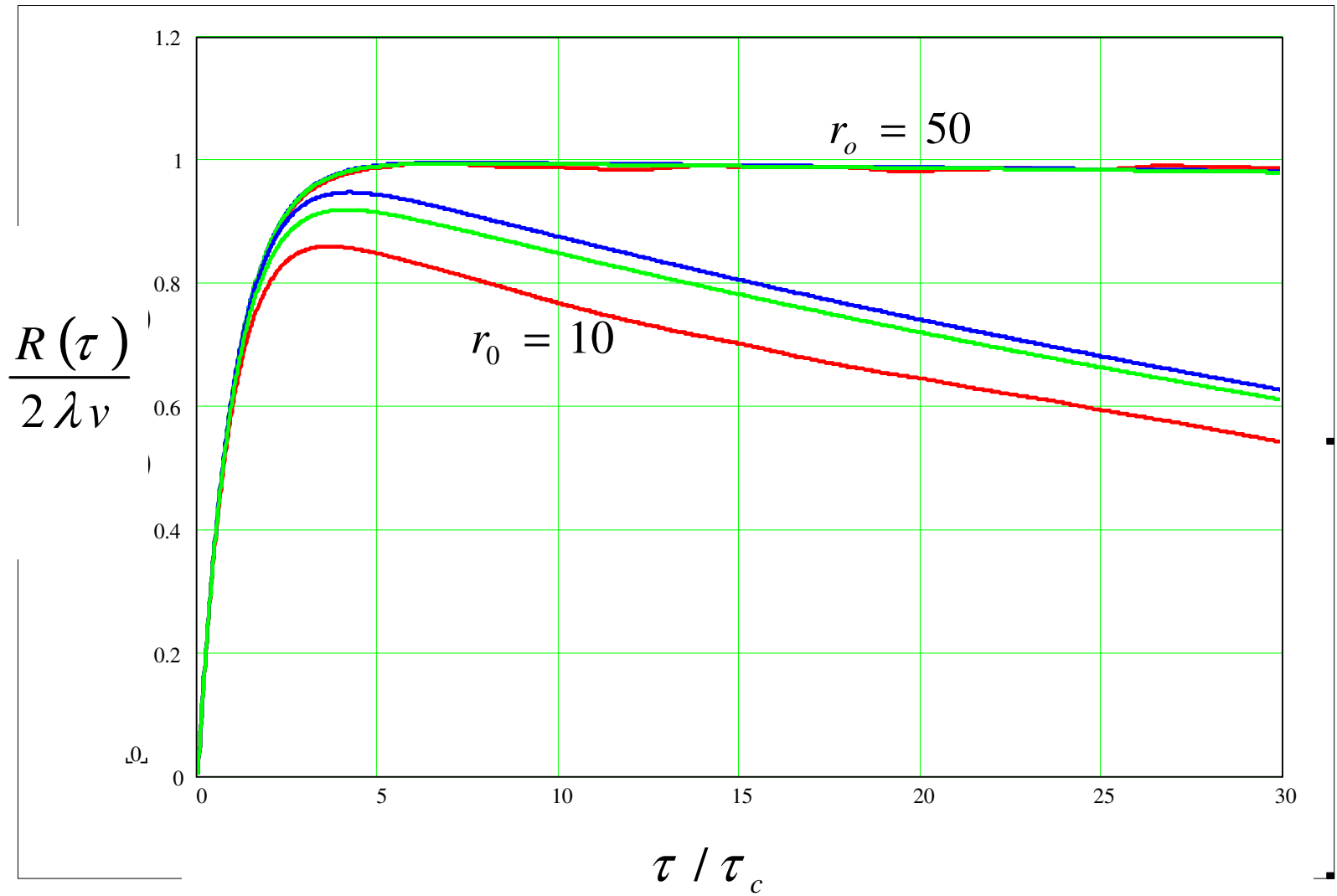
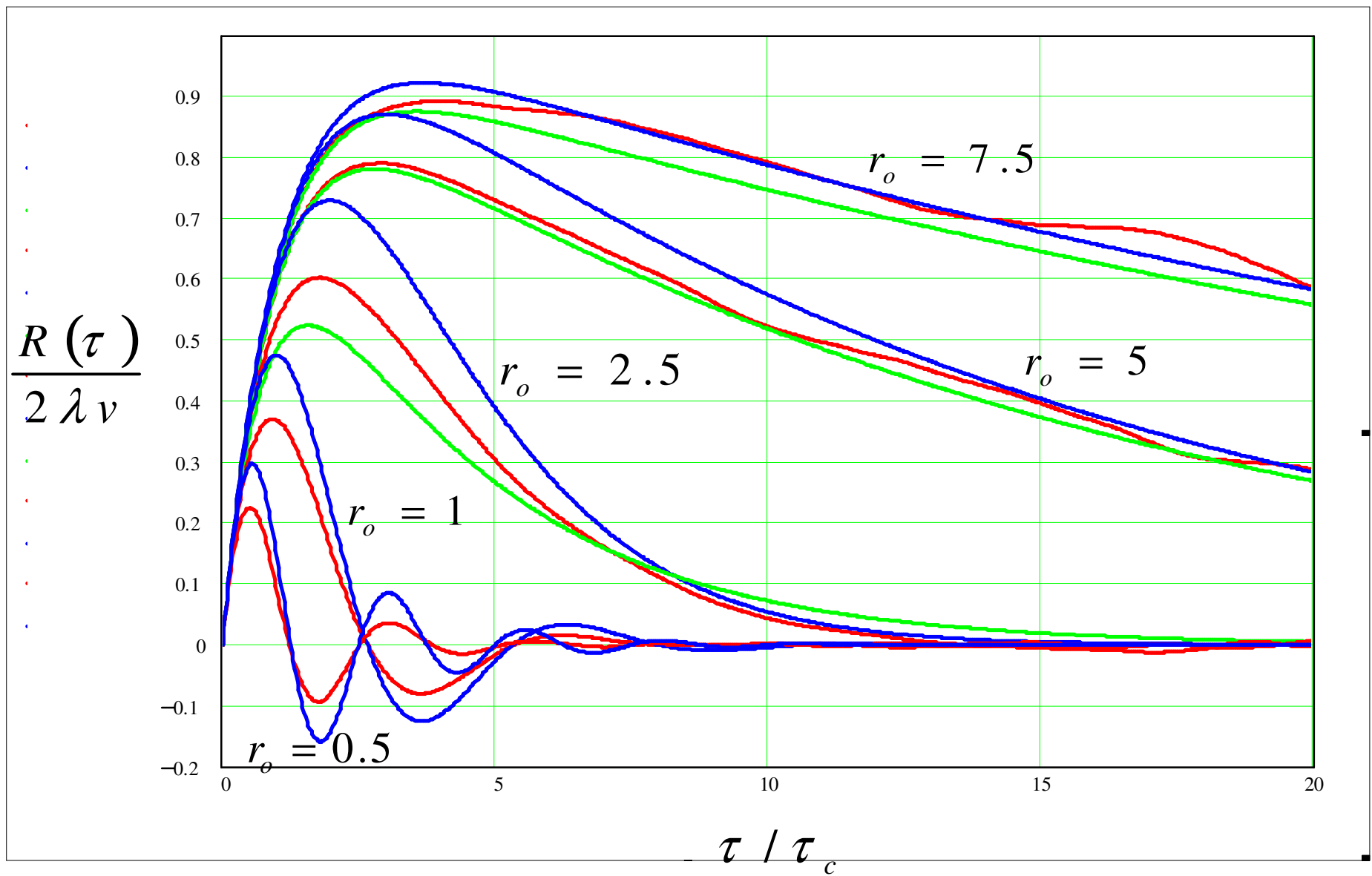
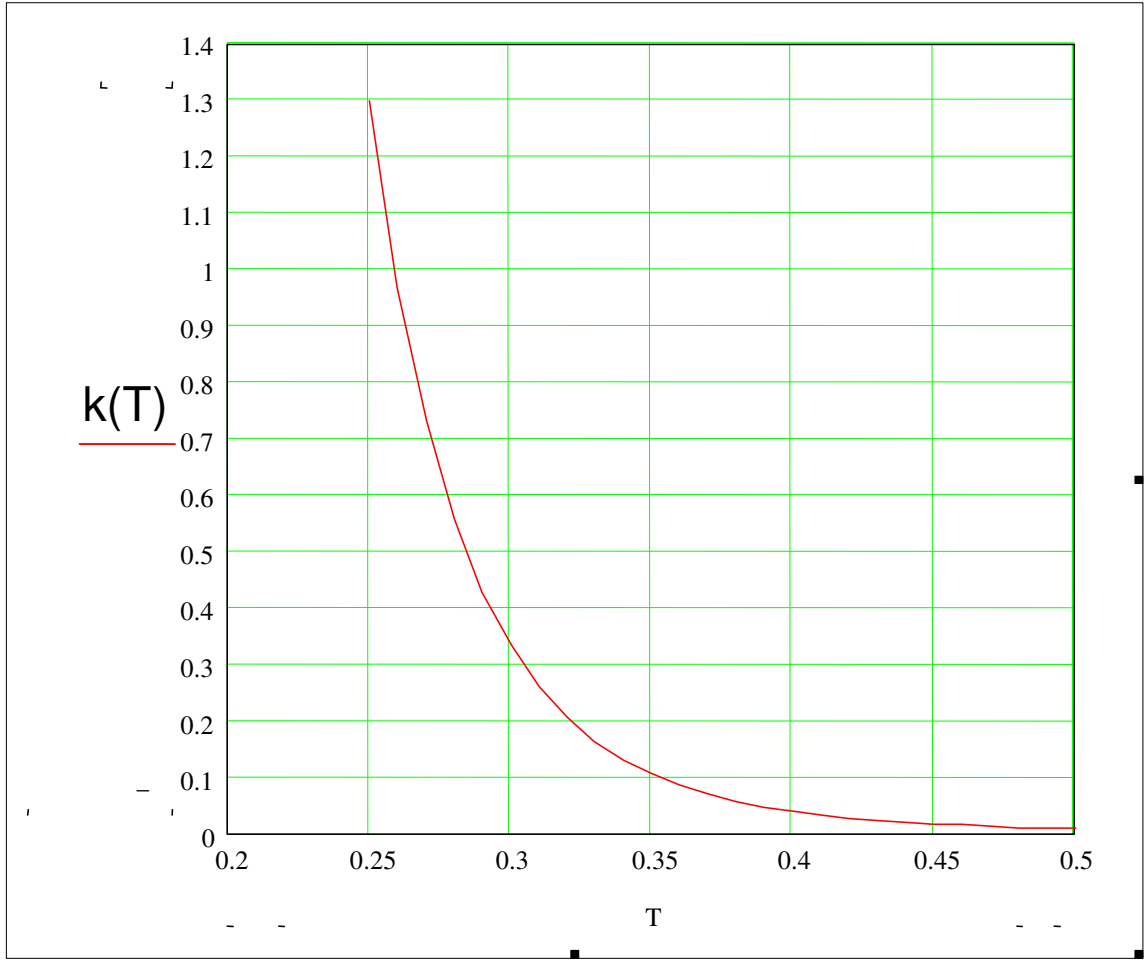
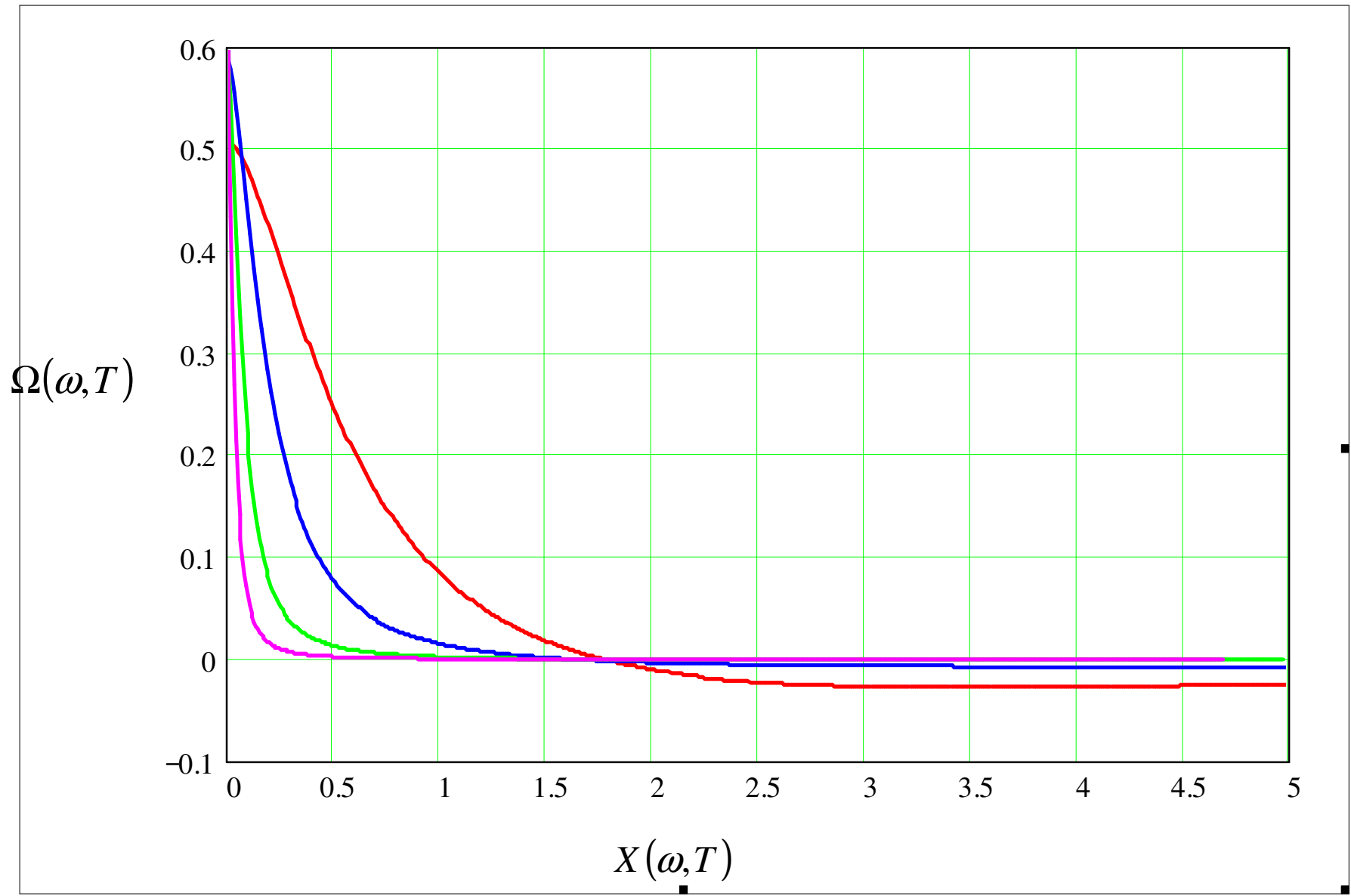
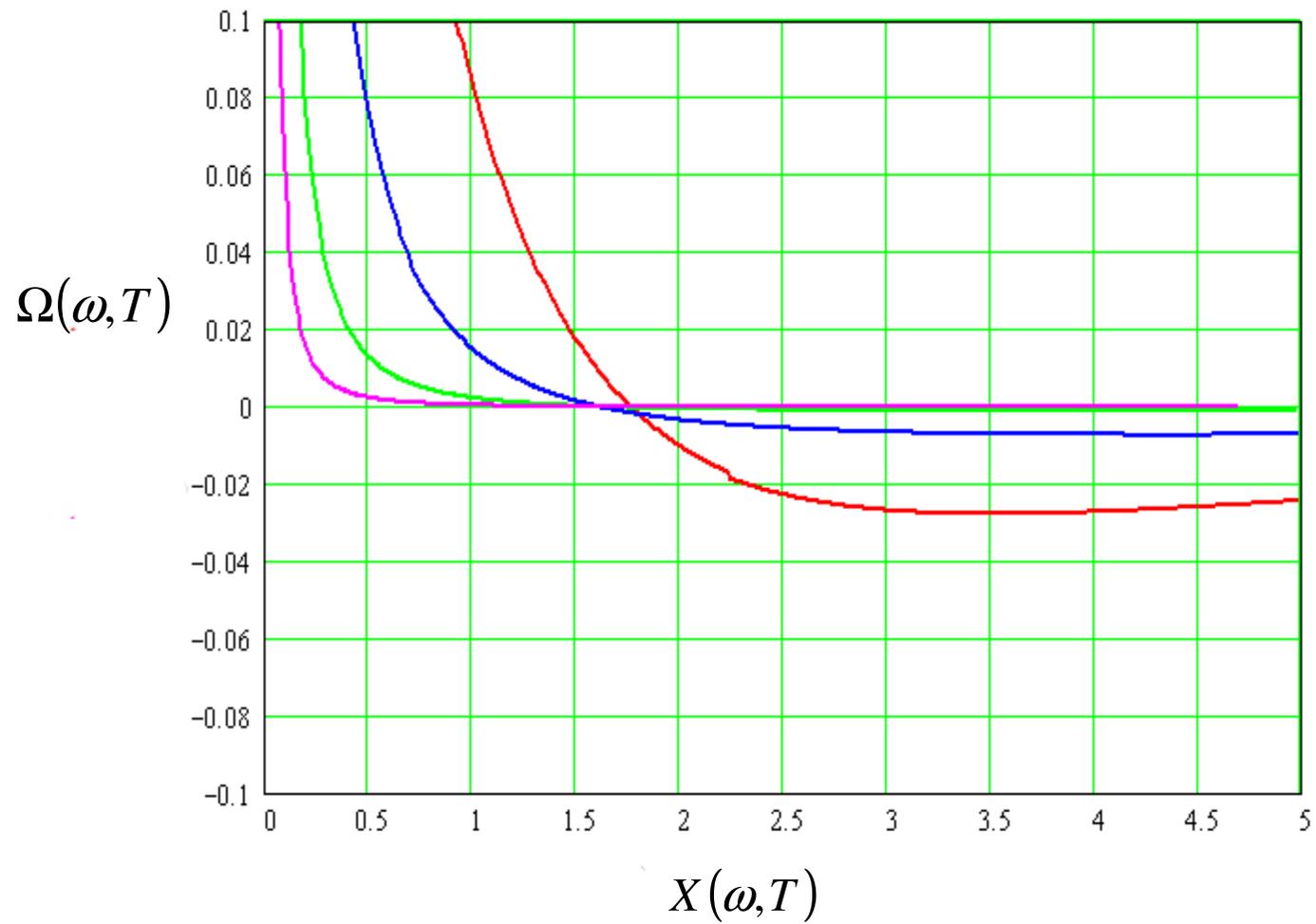
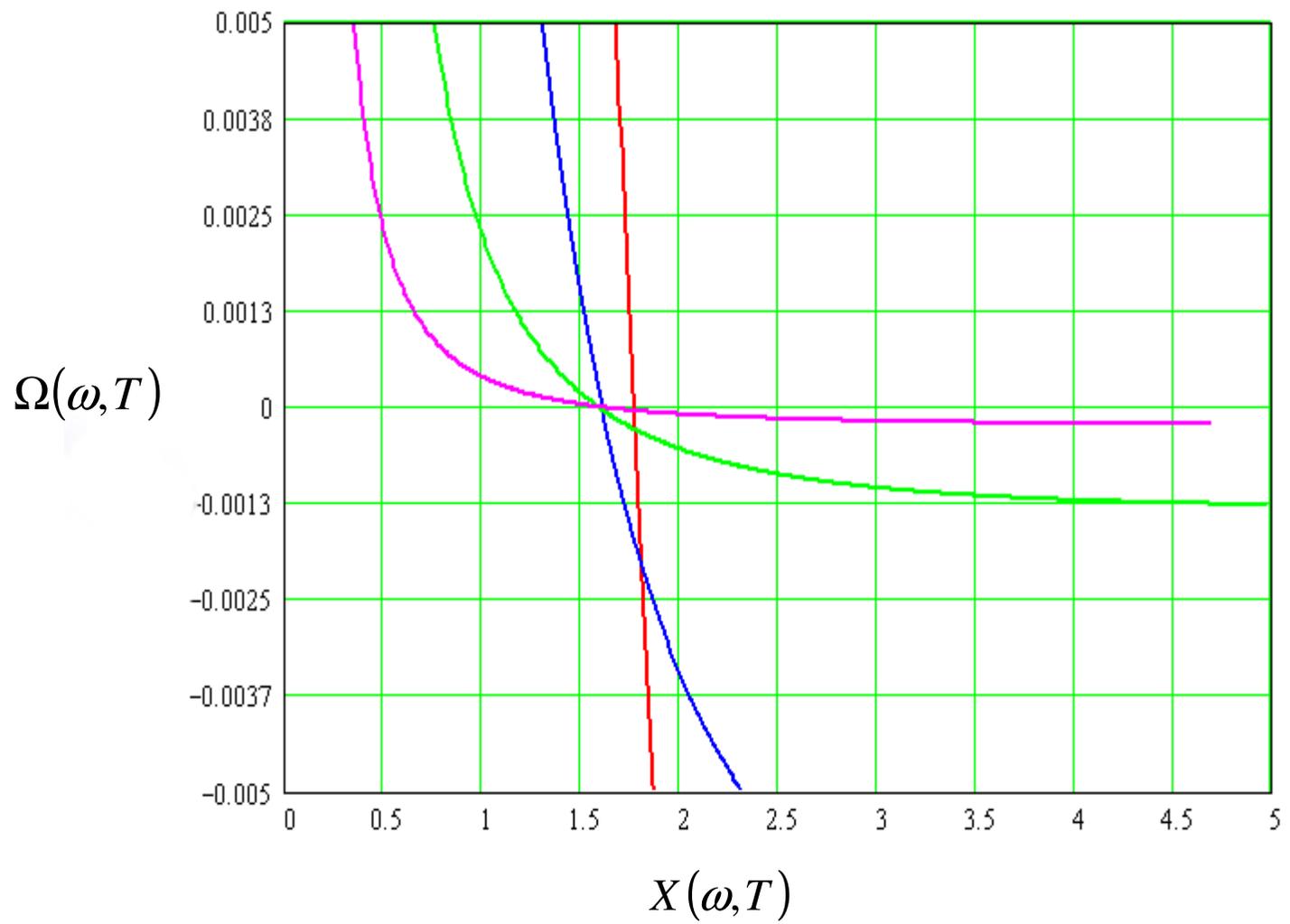
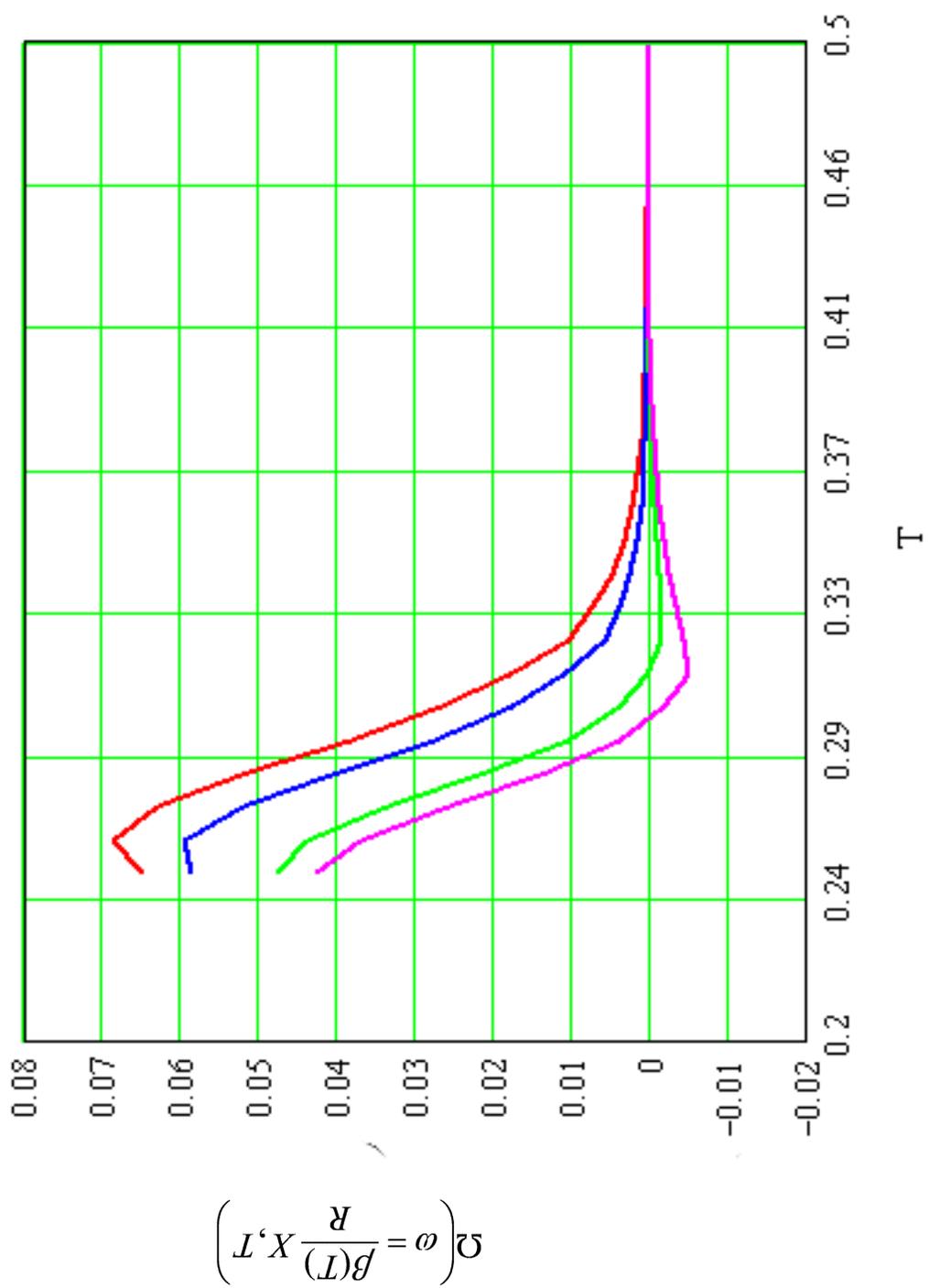
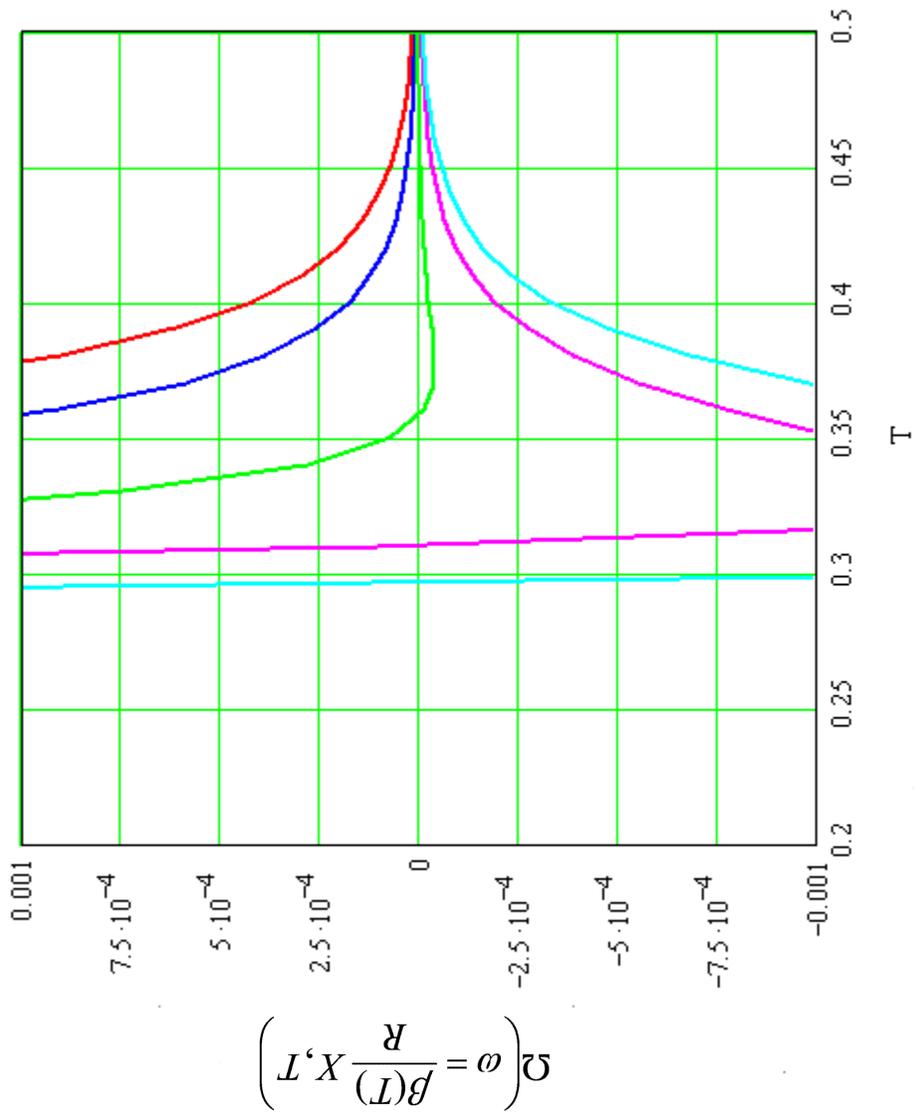
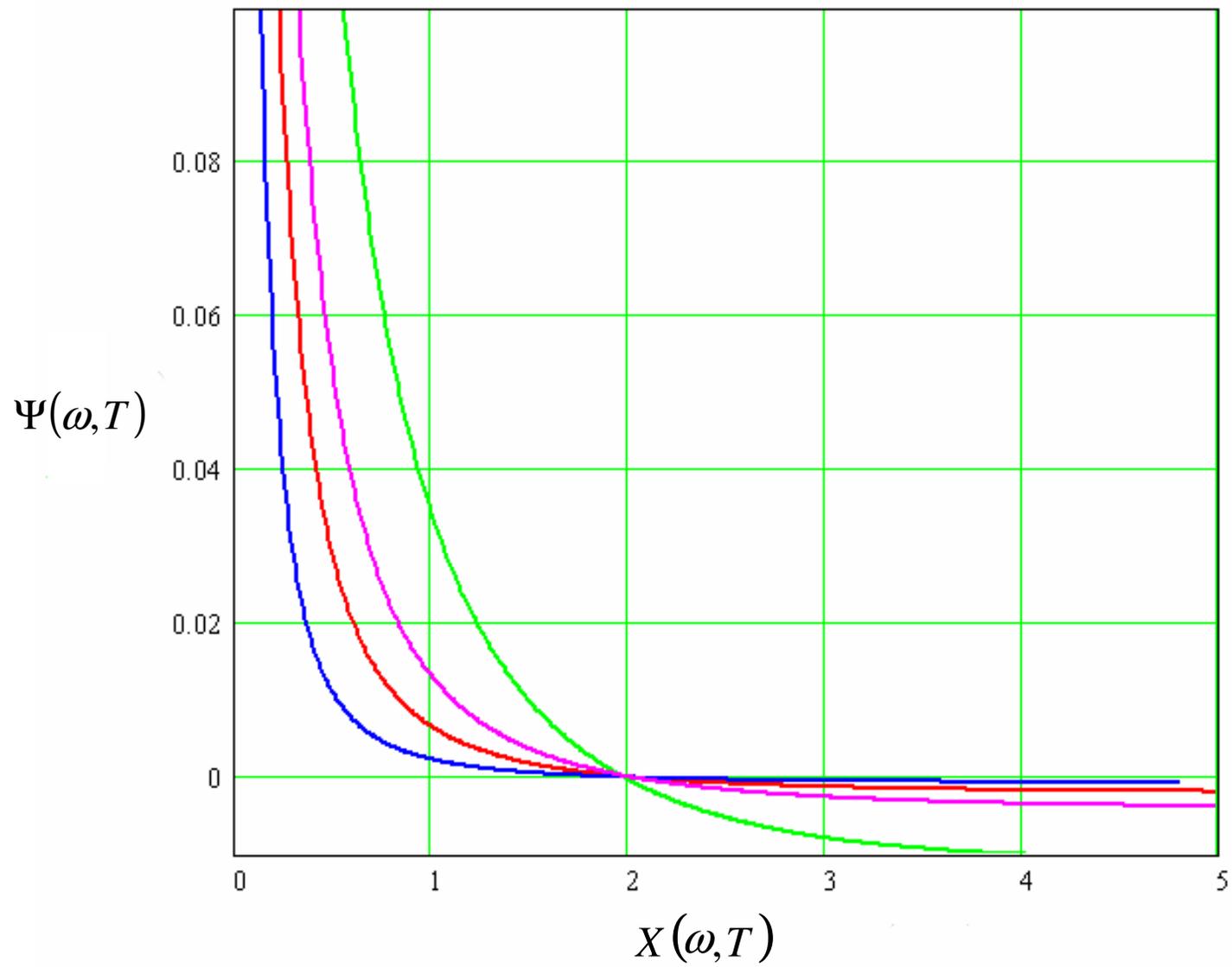
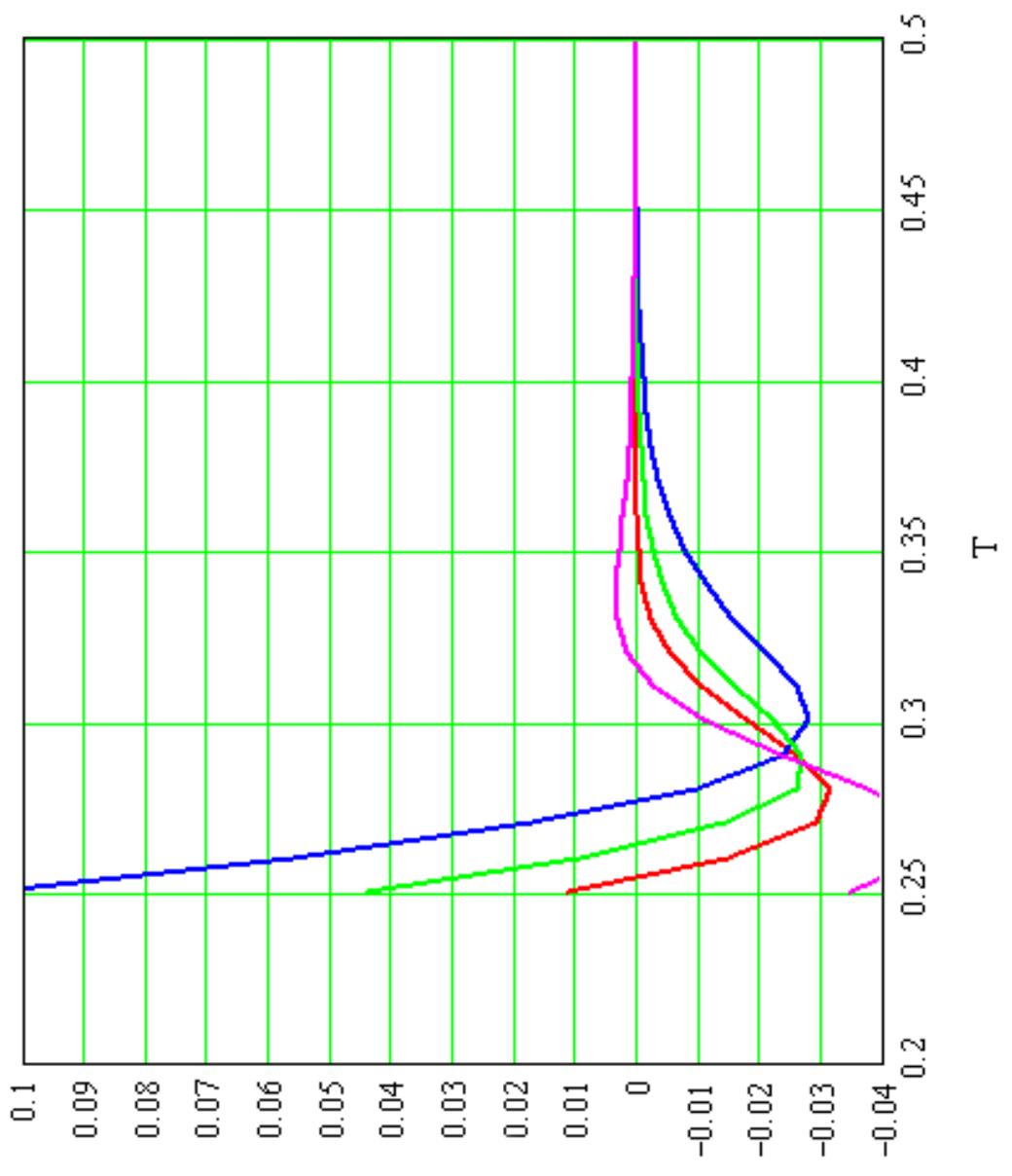
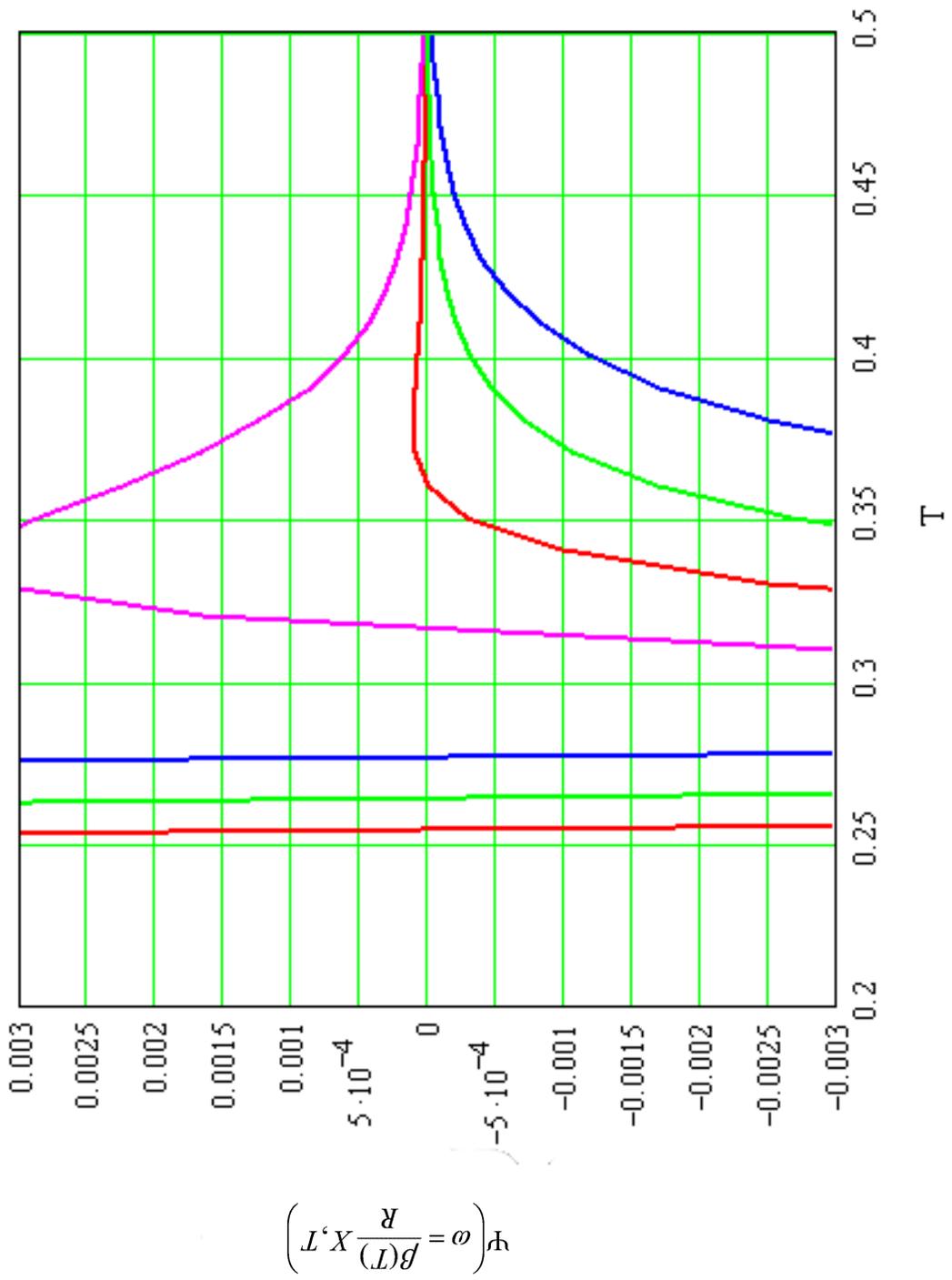
\end{document}